\documentclass[]{rspublic}

\usepackage[dvips]{graphicx}

\newcommand{\ea}{{\it et al. }}
\newcommand{\apj}{{\it Astrophys. J.}}
\newcommand{\aj}{{\it Astron. J.}}
\newcommand{\mnras}{{\it Mon. Not. R. Astron. Soc.}}
\newcommand{\aanda}{{\it Astron. Astrophys.}}
\newcommand{\ptra}{{\it Phil. Trans. R. Soc. A}}

\begin{document}

\title[Star cluster systems]{Massive star clusters in galaxies}

\author[W.~E.~Harris]{William E. Harris}
\affiliation{McMaster University, Hamilton ON L8S 4M1, Canada}

\label{firstpage}
\maketitle

\begin{abstract}{\bf globular clusters; stellar populations; galaxy
formation} The ensemble of all star clusters in a galaxy constitutes
its \emph{star cluster system}. In this review, the focus of the
discussion is on the ability of star clusters, particularly the
systems of old massive globular clusters (GCSs), to mark the early
evolutionary history of galaxies. I review current themes and key
findings in GCS research, and highlight some of the outstanding
questions that are emerging from recent work.
\end{abstract}

\section{Introduction}

The observational evidence available to date indicates that a major
star-forming epoch in a galaxy's history will generate a new set of
star clusters that accompanies its population of `field' stars. Thus,
it is physically meaningful to think of a \emph{subsystem} of star
clusters as consisting of all clusters formed in a given starburst,
and to treat the clusters as a proxy for the stellar subpopulation
formed in the same burst (see de Grijs 2010 and Larsen 2010 for more
extensive treatments of cluster formation in starburst
environments). The huge advantage offered by star clusters is that
they are easily bright enough to be measured individually within
galaxies as distant as 100 Mpc and even beyond, and in giant galaxies
particularly, they can be found in large numbers (see figure 1). We
can then construct \emph{distribution functions} of such key
parameters as mass, age and heavy-element abundance (metallicity) for
the clusters, instead of just the luminosity-weighted averages that we
get from the unresolved field-star light.

The Milky Way star cluster system (our starting point for all such
work and the `baseline' comparison system for all other galaxies)
separates out rather cleanly into the two classic subsystems: the
\emph{open clusters} (found throughout the disc and spiral arms along
low-eccentricity orbits) and the \emph{globular clusters} (GCs,
inhabiting the Galactic bulge and halo in a roughly isotropic
distribution of orbits). In addition, the GCs are distinctly older
than the open clusters (although with a small range of overlap around
$\sim 8$--10 Gyr), as well as more massive and less enriched in heavy
elements, indicating that they belonged to a brief early stage of
rapid star formation and chemical enrichment. The open clusters, like
the general population of disc field stars, are found at any age but
over a more restricted range of metallicities, marking the more
gradual ongoing growth of the Galaxy's disc.

But in even the nearest external galaxies (the Magellanic Clouds, M31
and the other Local Group galaxies), this convenient dichotomy
disappears. The Clouds, for example, contain small numbers of
classically old, massive, metal-poor GCs as well as many analogues of
open clusters, but we also find numerous examples of high-mass,
\emph{young} clusters that likely resemble GCs as they would have been
closer to their formation time. Investigations of star cluster systems
in other galaxies reveal still richer varieties, to the stage where
every part of the star cluster age/mass/metallicity three-parameter
space is occupied.

For thorough reviews and perspectives on the earlier literature up to
$\sim$2005, interested readers should see Harris \& Racine (1979),
Harris (1991, 2001), Ashman \& Zepf (1998) and Brodie \& Strader
(2006). The present article will concentrate on recent developments in
GC system (GCS) studies, leading up to a list of currently challenging
questions. This short and biased discussion unfortunately cannot do
justice to the diversity and richness of approaches now underway, and
(happily) it will be doomed to be quickly superseded by the rapid
advance of both theory and data. Perhaps the most important single
implication of the work in this area, however, is that the old GCs
represent a common thread in early galaxy evolution, dating back to
the first star formation within the pregalactic gas clouds.

\begin{figure}
\begin{center}
\vspace{-4.7cm}
\includegraphics[scale=0.55]{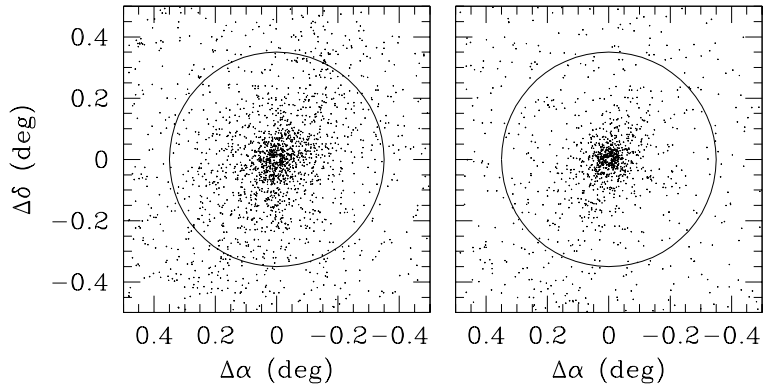}
\end{center}
\caption{A region near the centre of the Virgo giant elliptical galaxy
M87, showing the brightest of its total population of about 14\ 000
globular clusters (from Harris 2009$b$). {\it (left)} Distribution of
the blue, metal-poor GCs and {\it (right)} the more centrally
concentrated distribution of the red, metal-rich GCs. The circle in
each panel has a radius of 100 kpc, projected at the distance of the
Virgo cluster.}
\label{m87}
\end{figure}

\section{Data versus theory: some perspective}

Some perspective should be offered at this point about the links
between models and observations. This field is at basis a branch of
old stellar populations extended to the full range of galaxy types.
GCS studies began as a strongly data-dominated subject, with most
discoveries coming directly from photometric and spectroscopic surveys
that were guided primarily by the astrophysical intuition of the
observers. Quantitative models tended to follow later. Some branches
of astrophysics (e.g., Big-Bang cosmology, stellar structure, stellar
dynamics) have clear mathematical foundations accompanied by small
numbers of distinctive `parameters' that can be invoked to design new
observing programmes. This near-equal cooperation between models and
data is not currently the situation here, and a genuine understanding
of the formation and evolution of star cluster systems within their
parent galaxies is a considerably more complex issue. It starts on the
broad platform of the hierarchically merging galaxy-formation era, and
continues with the operation of gas dynamics at all scales, from
protogalaxies down to protostars, simultaneously with other key
elements including galaxy mergers, satellite accretions and dynamical
evolution of compact stellar systems. In the face of this complexity,
making the transition from models which are `scenarios' (even if
basically accurate) to ones that are both quantitative and realistic
in detail is a steep challenge. Nevertheless, with the rapid advance
of high-performance computation and the ability to simulate these
processes at a genuinely deep level, we can look forward to major
progress on the theory front.

At least some key features of GCS formation have, however, been
isolated. At one extreme lie the smallest dwarf galaxies that have
shallow potential wells which can support only a brief initial
starburst or else much slower, less dramatic star formation. These
early, low-metallicity \emph{pregalactic dwarfs} were probably the
formation sites of the metal-poor blue-sequence GCs (although also
leaving behind an intriguing `specific-frequency problem', see below)
(Searle \& Zinn 1978; Harris \& Pudritz 1994; Burgarella \ea 2001;
Beasley \ea 2002; Brodie \& Strader 2006; Mashchenko \ea 2008). At the
other extreme are the giant and supergiant ellipticals whose GCSs are
almost certainly composite populations. They will have GCs originating
from \emph{dissipational collapse} with huge amounts of gas undergoing
star formation spread over several episodes, \emph{dissipationless
accretion} of satellites coming in later with their own small
populations of GCs and \emph{major merging} of progenitor disc or
elliptical galaxies, with various mixtures of stars and gas. How
important each of these contributions will be for a given galaxy must
be driven by its size, environment and the stochastically governed
individual hierarchical-merging history. Major attempts to follow
these processes simultaneously have been carried out for giant
ellipticals in a semianalytic model (Beasley \ea 2003) and for the
Milky Way in cosmological simulations (Bekki \& Chiba 2001; Kravtsov
\& Gnedin 2005), but these are still exploratory and use prescriptive
rules for star and cluster formation. However, syntheses of these
approaches are developing (Bekki \ea 2008) and hydrodynamic
simulations of the important cases of GC formation in major mergers
and in pregalactic dwarfs are approaching the dynamical range required
to resolve individual clusters (Bournaud \ea 2008; Maschenko \ea
2008).

\section{Metallicity distributions}

A key and (fortunately) observationally prominent marker of evolution
in a stellar population is its heavy-element abundance or
\emph{metallicity}. The increase in mean metallicity with
time---driven by steady conversion of gas into stars and the cycle of
stellar evolution and supernova-driven enrichment---is extremely rapid
at the earliest times when the gas fractions are highest. It also
seems to be rapid in some of the most violent and shock-driven
star-forming environments. The \emph{metallicity distribution
function} (MDF) is thus one of the most important directly observable
tracers of this history.

\begin{figure}
\begin{center}
\includegraphics[scale=0.4]{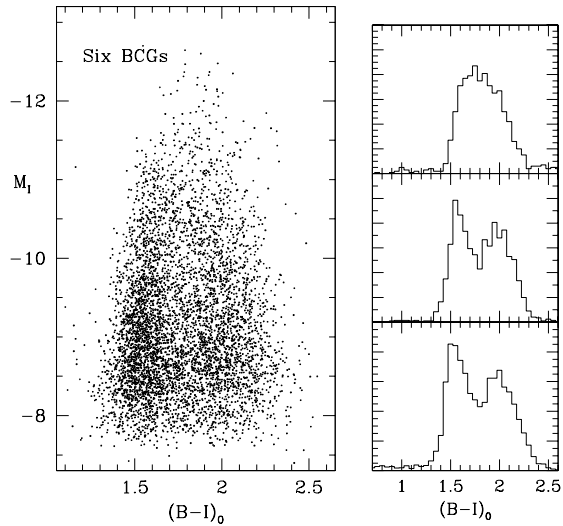}
\end{center}
\caption{\emph{(left)} Colour--magnitude distribution for a composite
sample of 5500 GCs in six giant elliptical galaxies (from Harris
2009$a$). Each point marks the total luminosity, $M_I$, and colour
index, $(B-I)_0$, of one cluster. Note the clearly defined blue
(metal-poor) and red (metal-rich) sequences. \emph{(right)} Histograms
of the colour indices subdivided into three luminosity zones, $M_I <
-10.4$ mag (top), $-10.4 < M_I < -9.4$ mag (middle) and $-9.4 < M_I <
-8.4$ mag (bottom). Note the clear bimodality for the lower-luminosity
regime, while in the highest-luminosity range the two modes overlap
more strongly.}
\label{cmd_6gal}
\end{figure}

A major discovery characterizing GCSs is that the whole MDF breaks
into a strikingly \emph{bimodal} form, with a metal-poor mode centred
near [Fe/H](MP) $\simeq -1.5$ dex and a metal-rich mode centred near
[Fe/H](MR) $\simeq -0.4$ dex. A graphic illustration of this
two-subpopulation division, as it is most frequently measured, is
shown in figure \ref{cmd_6gal}. Broadband photometric indices such as
in figure~2 for GCs older than $\sim 5$ Gyr become quite insensitive
to age and thus the colour reflects the cluster heavy-element
abundance surprisingly well. By convention, the metal-poor GCs are
often called the \emph{blue sequence} while the metal-richer ones are
the \emph{red sequence}. The translation from GC colour to metallicity
(which is monotonic, but may be nonlinear) is ultimately established
from spectroscopic abundance measurements, which have now been done in
several galaxies (see the work by Brodie \& Huchra 1990; Cohen \ea
1998, 2003; Kissler--Patig \ea 1998; Barmby \ea 2000; Perrett \ea
2002; Puzia \ea 2005; Strader \ea 2007; Cenarro \ea 2007; Woodley \ea
2009; and the fundamental data for the Milky Way clusters).

These two major GC sequences were first identified definitively in the
Milky Way with the work of Zinn (1985), who demonstrated that the two
subpopulations had distinct \emph{kinematic systemics and spatial
distributions} as well. The metal-poor cluster subsystem occupies a
more extended spatial distribution through the bulge and halo and is
kinematically more like an isotropic orbital distribution, while the
metal-rich subsystem defines a more centrally concentrated spatial
distribution and kinematically contains a higher component of ordered,
rotational motion somewhat resembling the field stars in the Galactic
`thick disc' or bulge.

Dozens of observational papers establishing the bimodal MDFs in other
galaxies exist, but key examples include Zepf \& Ashman (1993), Zepf
\ea (1995), Whitmore \ea (1995), Geisler \ea (1996), Gebhardt \&
Kissler--Patig (1999), Neilsen \& Tsvetanov (1999), Larsen \ea (2001),
Kundu \& Whitmore (2001), Peng et al. (2006) and Harris (2009$a,b$).
The factor-of-ten difference in heavy-element abundance between an
average metal-poor and a metal-rich GC becomes more sharply defined if
more metallicity-sensitive colour indices are used (cf. Harris \ea
1992; Geisler \ea 1996; Harris \ea 2004; Rhode \ea 2005; Peng \ea
2006; Spitler \ea 2008$a$; Harris 2009$a,b$). Direct spectroscopic
abundance measurements are best, although these entail far costlier
and more resource-intensive data acquisition. But the clear advantage
of photometric samples is that reliable first-order MDFs can be
readily obtained for large numbers of galaxies, offering us the
ability to track statistical trends across all types of galaxies and
environments. Bimodality is now seen so widely that any clear
deviations from it within an old GCS would now be regarded as
anomalous.

If bimodality is the major \emph{first-order} feature of the MDF,
there are now three \emph{second-order} trends providing intriguing
additional links to the early enrichment histories of both the host
galaxies and the clusters themselves:

\begin{enumerate} 
\item {\it Mean metallicity and dispersion versus galaxy size.} The
blue and red sequences individually appear at much the same mean
metallicity in all galaxies from giants down to dwarfs, and from
spirals to ellipticals. In addition, the intrinsic \emph{dispersion}
of each sequence (that is, the rms range in the cluster-to-cluster
differences in metallicity) is roughly constant at
$\sigma$[Fe/H](blue) $\simeq 0.30$ dex and $\sigma$[Fe/H](red) $\simeq
0.45$ dex. The intrinsic dispersion must represent the overall growth
of enrichment during a major star-forming period: a cluster forming
slightly later in the sequence will start with a larger amount of
pre-enrichment in its parent giant molecular cloud. However, a more
subtle trend is that the mean metallicity of each sequence increases
steadily with host-galaxy luminosity. Expressed in terms of
heavy-element abundance $Z$, the scaling relation is $Z \sim L^{0.2\pm
0.05}$ (Forbes \ea 1997; Larsen \ea 2001; Strader \ea 2004; Brodie \&
Strader 2006; Peng \ea 2006).

In qualitative terms, this mean-metallicity scaling indicates that a
random GC (either metal-poor or metal-rich) drawn from a
\emph{present-day} dwarf galaxy has a lower enrichment than one drawn
from a giant, thus arguing against pure major-merger or
dissipationless accretion scenarios in which (for example) the
blue-sequence clusters now present within a giant galaxy all formed
within smaller progenitors later accreted into the bigger central
potential. A \emph{caveat} to this view (see Brodie \& Strader 2006)
is that an isolated pregalactic dwarf should have a different and more
truncated enrichment history than its counterparts lying within the
deeper large-scale potential well of an emerging giant elliptical.
The key hint from the data, however, is that the \emph{global}
environment of the galaxy (on scales of $\sim 100$ kpc) has influenced
the \emph{local} formation conditions of its star clusters (on scales
of a few pc).

\begin{figure}
\begin{center}
\includegraphics[scale=0.4]{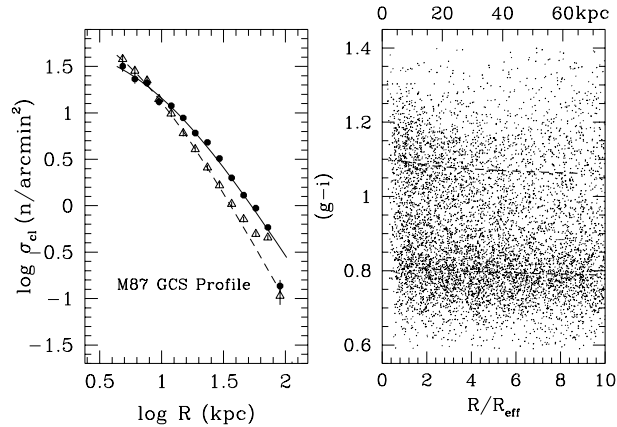}
\end{center}
\caption{\emph{(left)} Radial profile for the GCS in the Virgo giant
elliptical M87 (Harris 2009$b$). The metal-poor and metal-rich
clusters are shown as the solid dots and open triangles,
respectively. The red clusters form a more centrally concentrated
subsystem of the galaxy spheroid. \emph{(right)} Distribution of GC
colour $(B-I)_0$ versus galactocentric distance for the luminous GCs
in six supergiant ellipticals. The best-fit power-law curves for the
red and blue sequences are shown as the dashed lines, corresponding to
a metallicity scaling of $Z \sim R^{-0.1}$.}
\label{m87rad}
\end{figure}

\item {\it The mass/metallicity relation (MMR).} An unexpected finding
of new high-precision photometric surveys of GCSs in giant galaxies is
a correlation between the mean colours of the blue-sequence GCs and
their luminosities (e.g., Harris \ea 2006; Strader \ea 2006; Mieske
\ea 2006; Spitler \ea 2006; Forte \ea 2007; Harris 2009$a,b$;
Cockcroft \ea 2009; Peng \ea 2009). Effectively, the blue GCs become
progressively more heavy-element enriched at higher cluster mass. The
exact form of the correlation was a matter of debate in the first
round of papers, but a consensus has been emerging around two
conclusions: (i) the blue sequence is essentially vertical for the
less massive clusters in the range $M \lesssim 5 \times 10^5$
M$_{\odot}$, that is, $Z \sim L^0$ for low-mass GCs. But at
progressively higher $L$ (extending up to the most luminous GCs known
at $10^7$ L$_{\odot}$), the blue sequence bends towards redder
colours, eventually following a slope near $Z \sim L^{0.3}$ at its top
end (Harris 2009$a,b$; Peng \ea 2009). (ii) No such trend has been
found along the \emph{red} sequence, which is indistinguishable from
vertical in all systems studied so far. Detecting this MMR clearly
requires internally precise photometry and very large GC samples
because clusters above the $\sim 10^6$ M$_{\odot}$ level (where the
MMR becomes most noticeable) are rare and the changes in mean colour
are not large (see again figure~\ref{cmd_6gal}).

The MMR is, at this point, thought to be the result of some form of
self-enrichment of massive clusters during their formation period
(Strader \& Smith 2008; Bailin \& Harris 2009). A sufficiently massive
and dense protocluster ($\sim 10^7$ M$_{\odot}$ of gas within a 1~pc
protocluster core) can provide a sufficiently deep potential well and
enough dense gas to hold back a high fraction of the enriched
supernova ejecta from the first round of massive stars within the
protocluster, which can then enrich the still-forming
lower-main-sequence stars. For this mechanism to work, the entire
starburst needs to take 20--30 Myr to complete. For the highest-mass
GCs, the net enrichment over and above its \emph{pre-}enriched
metallicity will be $\Delta Z \simeq 0.05$ Z$_{\odot}$, large enough
to double its initial $Z_\mathrm{init} \simeq 0.04$ Z$_{\odot}$ along
the blue sequence. But the lower-mass clusters cannot be significantly
self-enriched because the supernova heating of the surrounding
protocluster ends up ejecting almost all gas not converted to
stars. Interestingly, this model predicts that the phenomenon should
affect the metal-rich GC sequence too, but it would be almost
undetectable on the red sequence because the incremental $\Delta Z$ is
always small relative to their pre-enriched level, $Z_\mathrm{init}
\simeq 0.4$ Z$_{\odot}$.

\item {\it Metallicity gradients.} The mean metallicity of the entire
GCS in large galaxies has long been known to decrease outward in the
halo. However, most of this global gradient is actually caused by the
changing proportions of red versus blue clusters with changing
galactocentric distance, $R_\mathrm{gc}$, that is, it is the result of
a \emph{population gradient} (e.g., Zinn 1985; Harris \ea 1992;
Geisler \ea 1996; Rhode \& Zepf 2004; Harris \ea 2006). One of the
more well-defined but not atypical examples is shown in figure
\ref{m87rad} for the Virgo giant elliptical M87 (Tamura \ea 2006;
Harris 2009$b$). Notably, the metal-rich GCs follow a spatial
distribution that mimics fairly closely the similarly metal-rich
spheroid light of the host galaxy, suggesting that they formed
together. The same view is supported by the details of the MDF for
both clusters and field-halo stars (see below). By contrast, the
shallower spatial distribution of the blue GCs more closely resembles
that of the isothermal dark-matter potential well, or is intermediate
between the dark matter and the halo light, consistent with the
interpretation that the blue GCs formed at quite an early stage.

But at a finer level, true metallicity gradients can be found
\emph{within} each of the two subpopulations. That is, the red and
blue GCs show an intrinsic decrease in mean metallicity with
increasing $R_\mathrm{gc}$, scaling as $Z \sim
R_\mathrm{gc}^{-(0.1-0.2)}$ (see figure~3b and Geisler \ea 1996; Lee
\ea 2008; Harris 2009$a,b$). In the hierarchical-merging picture, the
mean GC metallicity would reflect the depth of the larger-scale
potential well within which it was formed, being able to reach a
higher level of pre-enrichment deeper in to galaxy centre. Later
accretions of dwarf satellites, which have more metal-poor GCs, would
add to the population predominantly in the outskirts of the halo. A
potential but important complication is that later major mergers after
formation may have diluted the initial metallicity gradient (or, if
the merger brought in a very large amount of gas, it may have helped
rebuild the gradient for the metal-richer clusters only).
\end{enumerate}

\section{Age distributions and galaxy formation}

The Milky Way and its nearby satellites (the Magellanic Clouds,
Sagittarius and Fornax) contain the only populations of GCs for which
truly fundamental age calibrations can be achieved, based on isochrone
fitting to the deep, unevolved main sequence and even the white-dwarf
sequence. Recent published work (e.g., Gratton \ea 2003; De Angeli \ea
2005; Hansen \ea 2007; Mar\'{\i}n--Franch \ea 2009; among many others)
indicates that the absolute age of the metal-poor clusters is near 13
Gyr and shows a cluster-to-cluster dispersion that may be as low as
$\pm0.5$ Gyr, confirming their classic status as among the first
stellar structures to have formed in the Galaxy. The mean age for the
metal-richer subsystem is perhaps 2 Gyr younger than the extremely
old, metal-poor subsystem, and exhibits a higher cluster-to-cluster
age scatter, near $\pm 2$ Gyr rms.

\begin{figure}
\begin{center}
\includegraphics[scale=0.4]{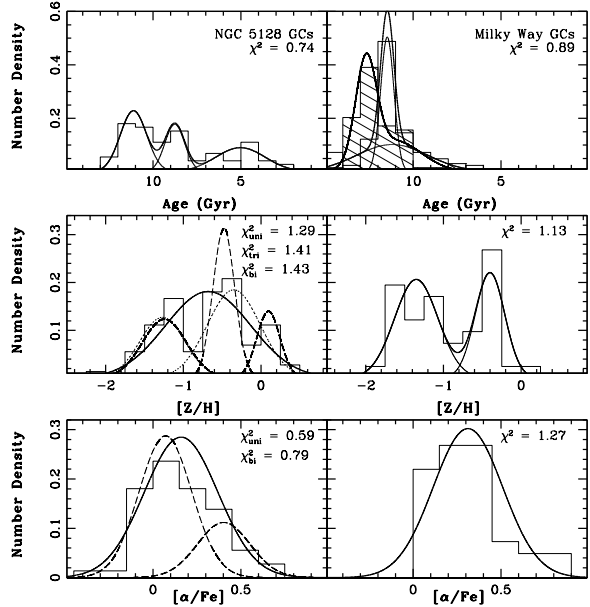}
\end{center}
\caption{\emph{(left)} Results for spectroscopically determined ages
and abundances of the GCs in the nearby giant elliptical galaxy NGC
5128. \emph{(right)} Ages and abundances for Milky Way GCs, determined
with the same techniques. The fitted lines show various unimodal and
bimodal Gaussian solutions to the age and abundance distributions in
each galaxy. (All data from Woodley \ea 2009).}
\label{ages_2gal}
\end{figure}

Direct age measurements for GCs in other, more distant systems are
extremely important but inevitably less precise. A method of attack
that has gained some valuable ground in the past decade has been the
use of spectroscopic line-strength parameters from the integrated
light of GCs, which are sensitive to different combinations of age,
$\tau$, metallicity, [Fe/H], and $\alpha$-abundance ratio,
[$\alpha$/Fe]. Because the Balmer lines are affected strongly by the
faint but numerous stars near the main-sequence turnoff, they are
sensitive to age, while the many different metal lines (Fe, Ca, Mg,
etc.) are more exclusively influenced by metallicity. From high
signal-to-noise ratio spectra at moderate resolution, age
determinations precise to $\pm 2$ Gyr (for the `old' age regime, $\tau
\sim 10$ Gyr) can be achieved. Key case studies can be found in
Proctor \ea (2004), Puzia \ea (2005), Pierce \ea (2006), Beasley \ea
(2008) and Woodley \ea (2009), among several others. A recent sample
for NGC 5128 is shown in figure \ref{ages_2gal}. Direct comparison
with the Milky Way clusers indicates that the NGC 5128 GCs exhibit a
comparable $\alpha$-abundance distribution, along with the familiar
bimodal MDF. However, the age distribution for the NGC 5128 GCS
clearly extends to younger ages, with a hint of major starbursts
having occurred at ages of $\sim 11$, 9 and 5 Gyr. By contrast, on the
same age-calibration scale, the Milky Way GCs range exclusively from
10 to 13 Gyr. Roughly similar age patterns show up in other large
ellipticals (see the references cited above). Although the
\emph{absolute} age scale for these spectroscopic-index programmes is
still a work in progress, the data so far strongly confirm that large
ellipticals are composite systems which assembled in steps from very
high redshift down to $z \sim 1-2$.

\section{Cluster sizes and the Fundamental Plane}

A surprisingly effective outcome of the {\sl Hubble Space Telescope
(HST)} imaging programmes for nearby galaxies and their GCSs has been
the ability to measure accurately the \emph{scale sizes} of the GCs,
i.e., their effective or half-light radii, $r_\mathrm{eff}$ or
$r_\mathrm{h}$. This is a dynamically valuable quantity because it
stays nearly invariant over many internal relaxation times and thus is
one of the best signatures we have of the linear size of the cluster
shortly after it formed its stars and ejected the remaining
protocluster gas. For galaxies within the Local Group, the structure
of the cluster can be resolved to within its core radius,
$r_\mathrm{c}$, (Barmby \ea 2007) and rather complete profile fitting
is readily possible. At increasing distances, the linear size of a GC
gradually shrinks relative to the $0.1''$ {\sl HST} resolution, but
the crucial scale radius $r_\mathrm{h}$ can be accurately measured for
galaxies out to 50 Mpc (major databases of this type are in Jord\'an
\ea~2005; Harris 2009$a$) by appropriate convolution with the stellar
point-spread function.

\begin{figure}
\begin{center}
\includegraphics[scale=0.4]{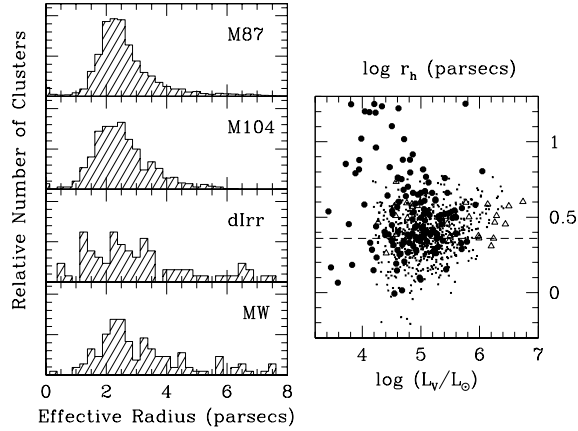}
\end{center}
\caption{\emph{(left)} Effective (half-light) radii, $r_\mathrm{h}$,
for old GCs in four systems: the giant elliptical M87, the giant Sa
M104, a combination of nearby dwarf galaxies and the Milky Way.
\emph{(right)} $\log r_\mathrm{h}$ versus cluster luminosity for GCs
in three massive disc galaxies: the Milky Way (large filled circles),
M31 (triangles) and M104 (small crosses). The median size,
$r_\mathrm{h} = 2.3$ pc, is shown as the dashed line.}
\label{sizes}
\end{figure}

Four representative sets of GC scale-radius data are shown in figure
\ref{sizes}: GCs in the supergiant elliptical M87 (Madrid \ea 2009),
the giant Sa disc galaxy M104 (Harris \ea 2009), a compendium of GCs
in nearby dwarf ellipticals (dEs, Georgiev \ea 2008) and the Milky Way
clusters (data from Harris 1996). The fundamental similarities among
these very different galaxies are striking (see, for example, Jord\'an
\ea 2005, who explore their potential as a `standard ruler'). The
asymmetric tail to larger $r_\mathrm{h}$ is the most noticeable
second-order difference. Clusters with $r_\mathrm{h} \gtrsim 5$ pc are
found predominantly in the dwarf galaxies and (for the Milky Way)
among the lower-mass clusters in the outskirts of the halo, apparently
reflecting their gentler tidal-field environment from birth. Da Costa
\ea~(2009) suggest that the distribution may be bimodal, with a
secondary mode at $r_\mathrm{h} \ge 8$ pc found in the dwarfs. The
intrinsic shape of the size distribution has been suggested to result
ultimately from a stochastic range in star-formation efficiencies
starting from a rather narrow range of initial protocluster sizes
(Harris \ea~2009$a$), followed by a longer phase of dynamical
evolution which shapes both the present-day size and mass
distributions (e.g., Fall \& Rees 1977; Gieles \& Baumgardt 2008).

More quantitatively, GC structural parameters have been used to define
a Fundamental Plane analogous to that for elliptical galaxies
(Djorgovski 1995). Their various structural properties are in fact so
tightly correlated that their structures are almost completely
specified by just two parameters, mass and (secondarily) central
concentration (McLaughlin 2000). A simple version of this manifold is
shown in figure \ref{sizes}b, where GC scale size is plotted versus
luminosity for three massive disc galaxies. GCs do not follow the
scaling relation $r_\mathrm{eff} \sim L^{1/2}$ that typifies only
slightly more massive, compact stellar systems such as dE nuclei and
ultracompact dwarf (UCD) galaxies (e.g., Evstigneeva \ea 2008; Mieske
\ea 2008). Instead, for GCs less massive than $\simeq 10^6$
M$_{\odot}$, we find $r_\mathrm{eff} \sim L^0$. A long-standing
question has been whether or not the GCs are truly distinct stellar
systems or if there is any kind of `bridge' connecting them to the
lower end of the dE/UCD manifold. The recent evidence, drawn from
galaxies with the richest GC populations in which we can find
significant numbers of high-mass clusters, is that the radii start to
increase for $M \ge 10^6$ M$_{\odot}$ and may merge fairly seamlessly
onto that upper manifold.

\section{The specific-frequency `problems' and formation issues}

Globular clusters that we see \emph{today} in the haloes of galaxies
of all types have typical masses of $10^5$ M$_{\odot}$, but the mass
of their original protocluster is likely to have been substantially
larger (and more so at higher mass). The raw star-formation efficiency
within the protocluster is expected to be $\sim30$\%. After formation
it undergoes continual mass loss dominated (at early times) by
supernovae and stellar winds from its massive stars, and at later
times by stellar evaporation and tidal truncation, reducing its total
mass after 12 Gyr by another factor of two or more depending on
initial mass (Baumgardt \& Makino 2003; Baumgardt \& Kroupa 2007;
McLaughlin \& Fall 2008). At a broader galaxy-wide level, the
high-resolution simulation of Bournaud \ea (2008) indicates that about
4\% of the gas in a relatively strong major merger is converted into
small, dense clouds that can be plausibly identified as protoglobular
clusters. Combining these arguments then suggests that present-day
galaxy haloes should have $\lesssim 0.5$\% of their stellar mass in
the GCS. For comparison, the most thorough observational estimates of
the mass fraction $M_\mathrm{GCS}/M_\mathrm{bary}$ (McLaughlin 1999;
Spitler \ea 2008$b$; Peng \ea 2008) place $M_\mathrm{GCS}$ in the
range 0.1--1\% of the total baryonic mass of a large galaxy,
correlating weakly with the luminosity of the galaxy itself and also
accounting for any nonstellar mass in X-ray halo gas. For dwarf
galaxies $\lesssim 10^8$ M$_{\odot}$, this mass fraction scatters much
more, ranging from zero to above 3\%. Spitler \& Forbes (2009) argue
that $M_\mathrm{GCS}$ may be most tightly determined by the total
dark-matter halo mass of the parent galaxy except, again, for the
smallest dwarfs.

\begin{figure}
\begin{center}
\includegraphics[scale=0.4]{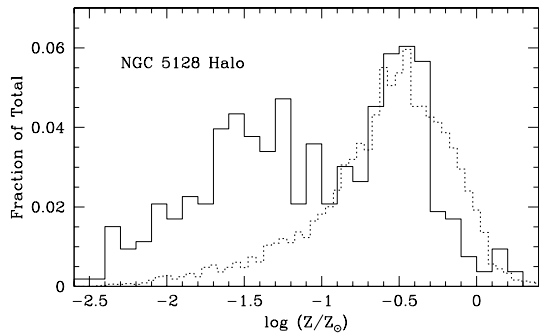}
\end{center}
\caption{Metallicity distribution function for the GCs in NGC 5128
(solid line; Harris \ea~2004 and Woodley \ea 2009) and for a sample of
halo field stars (dotted line; Rejkuba \ea 2005).}
\label{n5128halo}
\end{figure}

Recent simulations show that protoclusters of the right mass scale to
resemble GCs will form in both the low-metallicity environment of
pregalactic dwarfs (Mashchenko \ea 2008) and in gas-rich mergers of
large disc galaxies (Bournaud \ea 2008). But all of the contemporary
analyses lead to the conclusion that it takes a very large reservoir
of gas to make a typical GC. The actual conversion efficiency
(essentially, the ratio of GC protocluster mass to its local host
giant molecular cloud mass) may depend on merger shock velocities,
metallicities and other environmental factors, and needs to be much
better understood than it is now.

\emph{Specific frequency} $S_N$ is the number of GCs per unit galaxy
luminosity and is the most obvious observational proxy for GC
formation efficiency. Much of the original puzzle over the
order-of-magnitude differences in $S_N$ between (say) spirals and
ellipticals, or cD ellipticals versus field ellipticals (e.g., Harris
2001), has been alleviated by focusing on the mass ratio instead (see
above). The possibility that most of the metal-rich clusters in
elliptical galaxies formed through major mergers (Ashman \& Zepf 1992)
does not, by itself, solve the original `$S_N$ problem' because such
large amounts of incoming gas are needed to make the thousands of GCs
in giant ellipticals at any plausible formation efficiency that this
version of their origin becomes similar to basic hierarchical merging
(cf. Harris 2001). The $S_N$ values calculated separately for the red
and blue GCs vary systematically and nonlinearly with galaxy
luminosity, a pattern thoroughly discussed by Peng \ea (2008).

A newer `specific-frequency problem' is exemplified in figure
\ref{n5128halo}. The nearby giant elliptical system NGC 5128 offers a
rare opportunity to compare the MDFs of the GCs \emph{and} the halo
red-giant stars in the same galaxy. Our zeroth-order assumption is
that a major formation episode in a galaxy's history should produce
GCs in direct proportion to the amount of gas. If so, the MDFs of both
types of stellar populations should have the same shape, being merely
scaled versions of each other. This is plainly not the case (see
figure~6): there are about five times fewer metal-poor halo stars
\emph{per unit GC at the same metallicity} than there are metal-rich
halo stars.

The analyses for this and other galaxies (e.g., Forte \ea 2005) show
that metal-rich GCs match the mean metallicity and dispersion of the
halo giants well enough to further support the default interpretation
that they formed along with the bulk of the galaxy's spheroid. But
where are the many metal-poor stars that should go along with the
metal-poor clusters? Interestingly, similarly high $S_N$ values can be
found in some low-luminosity dEs, whose GCs are predominantly
metal-poor (Peng \ea 2008).

One possible solution is that the metal-poor halo stars are present,
but they occupy a much shallower spatial distribution than the
metal-rich spheroid, just as the low-metallicity clusters form a
spatially more extended subsystem than the metal-rich clusters. The
metal-poor field halo would become dominant only at larger
galactocentric distances than have normally been searched (see Harris
\ea 2007 for some tantalizing evidence along those lines). Another is
a timing argument: if the massive, dense protoclusters formed earliest
in the pregalactic clouds, and if later star formation were partially
truncated (perhaps by the epoch of reionization), then $S_N$(blue)
would end up artificially high (Harris \& Harris 2002). A step towards
a more quantitative interpretation has been developed by Peng \ea
(2008), who argue that the GC formation rate will scale with both the
star-formation rate and the star-formation-rate density, thus boosting
cluster formation at high redshifts. Much remains to be discovered
about the determining conditions of GC formation.

\section{Questions, puzzles and further directions}

The study of GC populations in galaxies provides a unique
observational window on galaxy formation. A shortlist of the questions
that have emerged from recent work would certainly include the
following:

\begin{enumerate}
\item What is the physical cause of bimodality? It does not
automatically emerge from current semianalytic or cosmological
simulations without deliberately inserting an external truncation
mechanism such as reionization (Beasley \ea 2002; Kravstov \& Gnedin
2005). Was the formation of intermediate-metallicity clusters
truncated near $z \sim 5$, \emph{in the same way in all major
galaxies?} Or were those clusters present but somehow biased towards
smaller masses, which would then have been removed more easily by
dynamical evolution?

\item Why was the formation of metal-poor, blue GCs so efficient
relative to the `normal', red GCs? Do the outermost parts of galaxy
haloes possess large numbers of metal-poor stars? More generally, what
features of the gas dynamics and composition inside a giant molecular
cloud determine the formation efficiency of massive clusters?

\item The formation of a giant elliptical galaxy may have finished
with a few major, gas-rich mergers, each of which generated metal-rich
clusters. Thus, the red GC `sequence' is, in itself, probably a
composite population. Is there a way to deconvolve the red sequence
into its age/metallicity components?

\item `Field' ellipticals typically have low specific frequencies and
are the most likely to have formed from late major mergers. But these
are very incompletely surveyed by comparison with the GC-rich
ellipticals in the Virgo, Fornax and Abell clusters. Do they show the
same features of bimodality, the same GC luminosity functions and the
same proportions of red/blue clusters as the rich ellipticals?

\item GCs more massive than $\sim 10^6$ M$_{\odot}$ belong to an
intriguing `transition region' characterized by an MMR, increasing
scale size, higher mass-to-light ratios and (as we see in massive
Milky Way GCs) multiple internal stellar populations. What drives
these systematic changes at the high-mass end of the GC sequence?
Cluster--cluster mergers? Increasing importance of dark matter? Can
all compact stellar systems, from GCs through UCDs and elliptical
galaxies, eventually be placed onto a single astrophysical sequence?

\item What is the total evolution of a massive star cluster from birth
to death? A comprehensive evolutionary model needs to include the gas
dynamics of its formation within a giant molecular cloud, early rapid
gas-mass loss, ongoing internal dynamical evolution, slower secular
evolution including the effects of the external tidal field and
eventual dissolution into the field. Pieces of these steps are
understood, sometimes in much detail, but a comprehensive end-to-end
story still awaits assembly.
\end{enumerate}

\label{lastpage}
\end{document}